\documentclass[12pt]{article}
   \newcommand{\beq}{\begin{equation}}
   \newcommand{\eeq}{\end{equation}}
   \newcommand{\beqa}{\begin{eqnarray}}
   \newcommand{\eeqa}{\end{eqnarray}}
   
   \newcommand{\dadb}[2]{\frac{{  d}#1}{{  d}#2}}

   \newcommand{\paraparb}[2]{\frac{\partial #1}{\partial #2}}
   \newcommand{\parsqaparbsq}[2]{\frac{\partial^2 #1}{\partial #2^2}}
   
   \newcommand{\bj}{{\mathbf j}}

   \newcommand{\bu}{{\mathbf u}}

   \newcommand{\bx}{{\mathbf x}}

   \newcommand{\bB}{{\mathbf B}}

 \newcommand{\Pc}{ {\mathcal{P}} }
 \newcommand{\bPc}{\mbox{\boldmath ${\Pc}$}}
 \newcommand{\Ic}{ {\mathcal{I}} }
 \newcommand{\bIc}{\mbox{\boldmath ${\Ic}$}}
 \newcommand{\paral}{\parallel }

\newcommand{\ed}{\end{document}}
\usepackage{color}
\usepackage{epsfig}
\usepackage{epstopdf}
\usepackage[a4paper,
            bindingoffset=0.2in,
            left=1in,
            right=1in,
            top=1in,
            bottom=1in,
            footskip=.25in]{geometry}
\usepackage{hyperref}
\hypersetup{
    colorlinks=true,
    linkcolor=red,
    filecolor=magenta,      
    urlcolor=cyan,
    citecolor = cyan,
    }
\begin{document}

\begin{center}
\noindent
{\large \bf A special exact class of three-dimensional equilibria \\ of a magnetized plasma exhibiting closed and nested \\ toroidal magnetic surfaces  \\ %\vspace{2mm} 
}\vspace{4mm}

D. A. Kaltsas$^{1,2}$, A. I. Kuiroukidis$^1$,  and G. N. Throumoulopoulos$^1$ \vspace{4mm}

$^1$Department of Physics, University of Ioannina, GR 451 10 Ioannina, Greece\\
$^2$Department of Informatics, Democritus University of Thrace, GR 654 04 Kavala, Greece  \vspace{4mm}

\vspace{3mm}
Emails: kaltsas.d.a@gmail.com, \  a.kuirouk@uoi.gr,\   gthroum@uoi.gr
\end{center}

%%%%%%%%%%%%%%%%%%%%%%%%%%%%%%%%%%%%%%%%%%%%%%%%%%%%%%%%%%%%%%%%%%%%%%%%%%%%%%%%%%%%%%%%%%%%%%%%%%%%

\begin{abstract}

We construct a special class of three-dimensional (3D)  equilibria with pressure anisotropy which showcase closed, nested toroidal magnetic surfaces  that are strongly asymmetric in the toroidal direction by applying a sinusoidal perturbation to the axisymmetric Solov'ev equilibrium. They also exhibit distinct closed and nested current-density surfaces.  For certain values of the free parameters involved, the perturbations lead to the formation of  magnetic islands and stochastic areas in the outer plasma region, while well-defined magnetic surfaces persist in the inner region. In addition, it is demonstrated that the existence of closed and nested surfaces within the plasma region on which the magnetic field modulus is uniform (isomagnetic surfaces)  is neither necessary nor sufficient for the existence of respective closed and nested magnetic surfaces.

\vspace{1mm}


\end{abstract}

%\newpage

%%%%%%%%%%%%%%%%%%%%%%%%%%%%%%%%%%%%%%%%%%%%%%%%%%%%%%%%%%%%%%%%%%%%%%%%%%%%%%%%%%%%%%%%%%%%%%%%%%%%

\section{Introduction}\
Owing to the mathematical difficulty,  a general theory concerning the
existence of solutions to the magnetohydrodynamic (MHD) equilibrium equations is not available  \cite{Su,SaYa}. In particular, 
magnetic confinement of laboratory fusion plasmas  involves  the existence of  equilibrium states with well defined closed and nested toroidal magnetic surfaces. 
For two-dimensional (2D) MHD equilibria, e.g. in the presence of axisymmetry, the existence of such surfaces is rigorously guaranteed. However,  in the absence of any continuous spatial symmetry the existence of steady states with everywhere smooth, nested toroidal magnetic surfaces  has been questioned because the breaking of symmetry allows for
magnetic field-line braiding \cite{Gr,St,Gr85}. There are several  papers with results compatible  with this conjecture, i.e. in \cite{Mo} it is demonstrated that the magnetostatic equilibria that are topologically accessible from a given initial magnetic field generally contain tangential discontinuities (i.e. current sheets),  even although the initial field may be infinitely differentiable. In \cite{BrLa} it was proved that 3D sharp (fixed) boundary solutions of the MHD equilibrium equations with toroidal magnetic surfaces exists when departure of axisymmetry is sufficiently small and the pressure is a step function.  
In addition, the non-existence of 3D isodynamic equilibria, i.e. equilibria with isomagnetic surfaces coinciding with the magnetic surfaces, has been proven in \cite{PaBa83,Pa86}; the only equilibrium of this kind is  axisymmetric \cite{Pa68}. The interest in isodynamic equilibria was motivated because in this case the guiding-center grad-$B$ drift perpendicular to the magnetic surfaces vanishes, thus potentially leading to reduced transport processes.

For these reasons, in order to obtain   steady states with favourable confinement properties, particularly in the framework of stellator optimization, some kinds of symmetry have been imposed, e.g. quasisymmetry in which $B$  having a continuous symmetry in certain coordinate systems, e.g. Boozer coordinates, becomes independent of one of the coordinates 
\cite{Hel,LaMe,Pl}.  
Necessary and sufficient conditions for a magnetic field to be quasisymmetric, though not necessarily pertaining to an MHD equilibrium,  was derived  in \cite{Ro}; the condition states that in order for $\bB$ to be quasisymmetric,  there should exist a vectorial field $\bu$ and a scalar function $\psi$ satisfying the set of equations 
$
\bu\cdot\nabla B=0\,, \ \bB\times \bu =\nabla \psi \ \mbox{and} \  \nabla \cdot \bu =0
$.
In addition to the existence of magnetic surfaces  labelled by $\psi$, these conditions imply the existence of distinct isomagnetic surfaces to which this field $\bu$ is tangential. However, they do not guarantee the closeness of either  the magnetic or  the isomagnetic surfaces. Also, in \cite{RoBa} it is argued that pressure anisotropy extends the regime of    quasisymmetric equilibria  beyond the isotropic pressure equilibria commonly considered. In this respect,  the problem of finding a three-dimensional solenoidal vector field such that both the vector field and its curl are tangential to a given family of toroidal surfaces was studied in \cite{SaYa} and smooth quasisymetric solutions foliated by toroidal surfaces that are not invariant under continuous Euclidean isometries were also constructed, and they were identified as equilibria with anisotropic pressure.

More recently, there are publications contradicting the conjecture of non existence, i.e. in \cite{CoPa} 3D solutions of the magnetohydrostatic  equations in bounded domains and on the torus   were constructed, as infinite time limits of Voigt approximations of viscous, non-resistive incompressible magnetohydrodynamics equations.  In \cite{CaDu} it is proved that an MHD equilibrium with non-constant pressure on a compact three-manifold with or without boundary admits no continuous Killing symmetries for an open and dense set of adapted metrics; this contrasts with the  conjecture in \cite{Gr,Gr85} in which is loosely stated that an MHD equilibrium on a toroidal Euclidean domain in $\mathcal{R}^3$ with pressure function foliating the domain with nested toroidal surfaces must admit Euclidean symmetries. 
In a spirit similar to the present study, weakly asymmetric plasma equilibria with nested toroidal magnetic surfaces and isotropic pressure were constructed by a method that expands up to first order about the axisymmetric configuration without any symmetry assumption \cite{SoIl,So,IvSo}. This indicates the existence of analytical  counterexamples  to the aforementioned conjecture of non-existence of such 3D equilibria. The construction, performed without the use of flux coordinates, is based on a generalization of the Grad-Shafranov (GS) equation to include magnetic-field asymmetry and by introducing weak sinusoidal perturbations in such a way that the perturbed equilibria satisfy the MHD equilibrium  equations. The method was exemplified by employing as unperturbed equilibrium the well-known and widely employed Solov'ev solution.

Aim of the present study is the construction of 3D toroidal equilibria with pressure anisotropy, exhibiting  closed and nested magnetic surfaces in a different setting. It is based on two foundations:
%ingredients:  
first, a special class of  equilibria  with anisotropic pressure elements given by Eqs. (\ref{sol}) valid for any (asymmetric) magnetic field, which was identified for the first time in \cite{LoSp}; and  second, by introducing the representation (\ref{Solp}) for the magnetic field in cylindrical coordinates ($r,\phi,z)$ involving three  arbitrary  functions, i.e. the functions $U(r,\phi,z)$ and $w(\phi)$ periodic in $\phi$ and the function $F(r,z)$. Then, these functions are specified  as sinusoidal  perturbations to the Solov'ev equilibrium to construct 3D steady states. Also, in addition to the impact of the free parameters on the equilibrium characteristics, the existence of isomagnetic surfaces and their properties are examined systematically. Compared with the asymmetric equilibria of \cite{SoIl,So,IvSo},  the present study is different in three respects: first, the representation (\ref{Solp}) for the magnetic field is inherently divergence free, while in \cite{SoIl,So,IvSo} the equation $\nabla \cdot \bB=0$ is included in the set of equations to be solved; second, the pressure in \cite{SoIl,So,IvSo} is isotropic, while here is anisotropic; and third, unlike the weak perturbations considered in \cite{SoIl,So,IvSo},  here the perturbations are arbitrarily strong. The most important result is the  construction of 3D strongly toroidally asymmetric  equilibria based on the exact spatial solution (\ref{sol}), showcasing  closed and nested magnetic surfaces. Also,  the existence of nested isomagnetic surfaces  is examined in conjunction  with their relation to the existence of closed and nested  magnetic surfaces.

The rest of the paper is organized as follows. In Sec. 2 the axisymmetric Solov'ev solution is reviewed. The  3D equilibrium equations with anisotropic pressure is presented in Sec. 3  together with the exact  solution and the inherently divergence free representation of the magnetic field.  Equilibrium solutions showcasing closed and nested magnetic surfaces with both weak and strong toroidal asymmetry is the subject of Sec. 4. Certain poloidal cross-section considerations are conducted in  Appendix A. Sec. 5 summarizes the conclusions. 

\section{Axisymmetric Solov'ev equilibrium}
\begin{figure}[h]
%\vspace{-0.4cm}
\begin{center}
\includegraphics[width=0.49\linewidth]{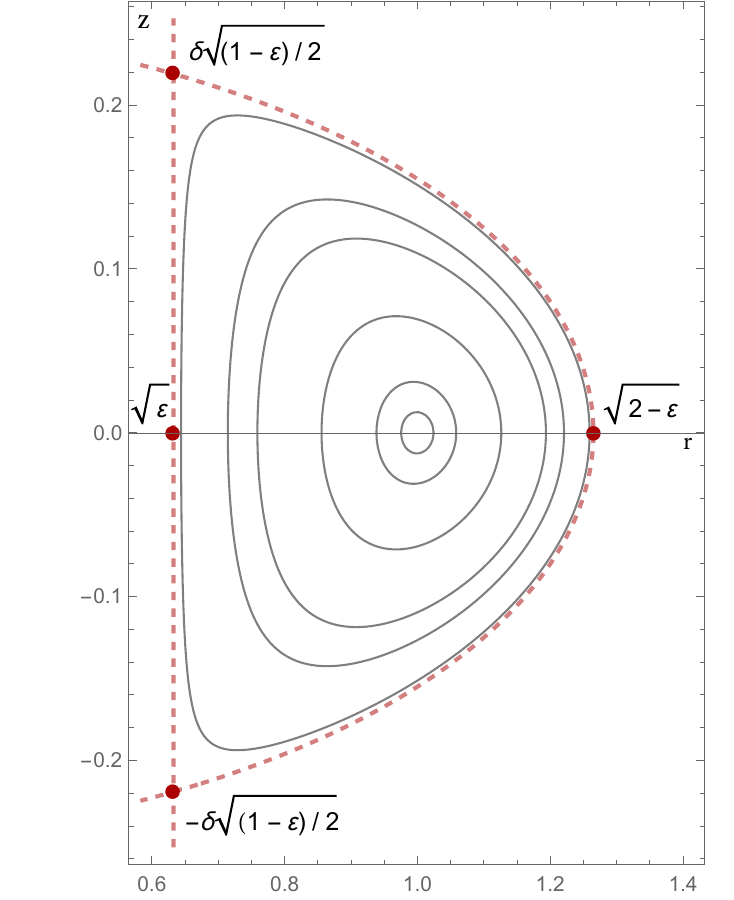}
\end{center}
\caption{The axisymmetric diamagnetic Solov'ev configuration for $\delta=\epsilon=0.4$.  }
                                                          \label{SolF}
\end{figure}

 We start from the GS equation in SI units which describes axisymmetric MHD equilibria with isotropic pressure:
\beq
\label{GS}
\Delta^\star \psi + F\dadb{F}{\psi} + \mu_0 r^2 \dadb{P}{\psi} = 0\,, \quad \Delta^\star := \parsqaparbsq{}{r} + \parsqaparbsq{}{z} -\frac{1}{r} \paraparb{}{r},
\eeq
where $\psi(r,z)$ is the poloidal magnetic flux function which labels the magnetic surfaces;  $P(\psi)$ and $F(\psi)$ are  the free functions of pressure and poloidal current. Introducing the normalized quantities $\hat{r}=r/r_a$, $\hat{z}=z/r_a$, $\hat{\psi}=\psi/(B_a R_a^2)$, $\hat{F}=F/(B_a r_a)$ and $\hat{P}=P/(B_a^2/\mu_0)$,  where $r_a$ and $B_a$ are a reference radial length and magnetic  field, respectively, (\ref{GS})  is written  in the dimensionless form (omitting hats) $\Delta^\star \psi + F dF/d\psi + r^2 dP/d\psi=0$. Choosing the free functions as $P(\psi)=P_a - 2(1+\delta^2)\psi$ and $F(\psi)=(F_0^2+ 4 \epsilon \psi)^{1/2}$, where $\delta$, $\epsilon$, $F_0$ and $P_a$ are free parameters, the resulting equation admits the Solov'ev solution \cite{Sol}
\beq
\label{Sol}
\psi_{ax}=z^2(r^2-\epsilon)+\frac{\delta^2}{4}(r^2-1)^2.
\eeq
 The respective up-down symmetric equilibrium is diamagnetic for $\epsilon\geq 0$ and paramagnetic for $\epsilon <0$. In the present study  we restrict ourselves to the diamagnetic configuration, shown in Fig. \ref{SolF}. 
It forms spontaneously a separatrix consisting of an  outer elliptic part and an inner part parallel to the axis of symmetry, thus having a couple of X-points.  The magnetic axis is located at ($z=0,\ r=1$) on which $\psi_{ax}=0$ and, therefore, the reference length $r_a$ corresponds to the radial distance of the magnetic axis. Since the plasma boundary is not imposed from the outset, it can be chosen a posteriori to coincide with the separatrix or with one of the internal, smoothly closed magnetic surfaces. In the former case the configuration's  triangularity is 1, the aspect ratio ($\sqrt{2-\epsilon}+\sqrt{\epsilon})/(\sqrt{2-\epsilon}-\sqrt{\epsilon})$ and  the elongation  $\delta \sqrt{2(2-\epsilon)}/(\sqrt{2-\epsilon}-\sqrt{\epsilon})$.  Thus, $\epsilon$ determines the compactness of the configuration, while the elongation depends mainly on $\delta$ with a weaker contribution from $\epsilon$. For $\epsilon>0$ the solution describes a tokamak-type equilibrium, while for $\epsilon=0$ reduces to a spheromak one. For $\epsilon=F_0=0$, the equilibrium reduces to the Hill's vortex with a purely poloidal magnetic field.  The toroidal component of the magnetic field is $B_\phi=(F_0^2+ 4 \epsilon \psi)^{1/2}/r$, with $F_0/r$ the vacuum contribution created by external currents, and the poloidal magnetic field is $\bB_p=\nabla\phi\times \nabla \psi_{ax}$. The respective current-density components are $j_\phi=\Delta^\star \psi_{ax}/r=2r(1+\delta^2)-2\epsilon/r$ and $\bj_p=-\nabla \phi\times \nabla F$. The pressure takes the maximum value $P_a$ on the magnetic axis such that it decreases monotonically and vanishes at the boundary.
 
\begin{figure}[h]
%\vspace{-0.4cm}
\begin{center}
\includegraphics[width=0.47\linewidth]{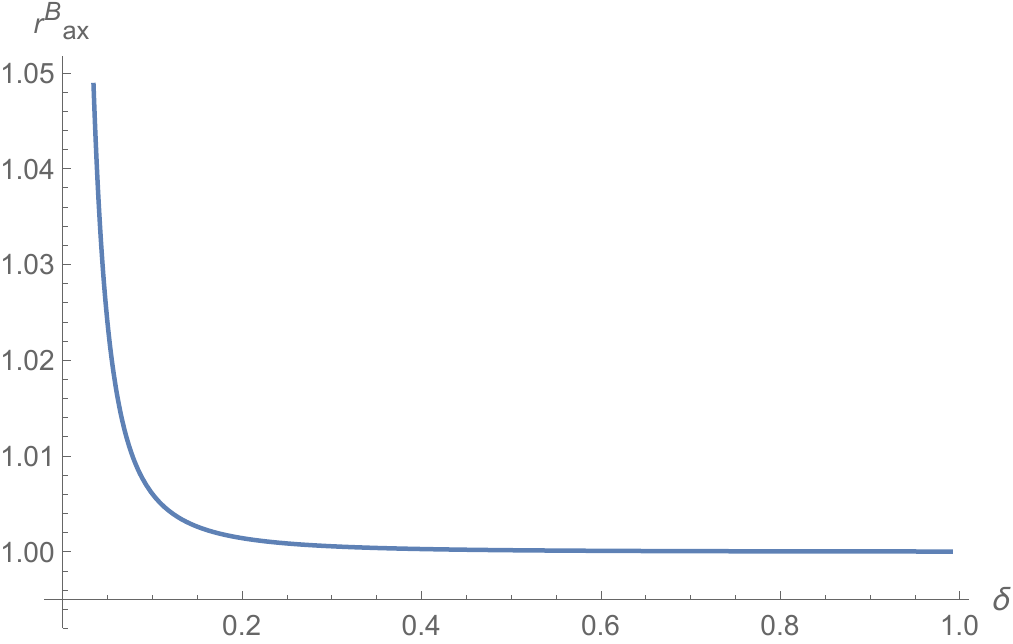}
\includegraphics[width=0.47\linewidth]{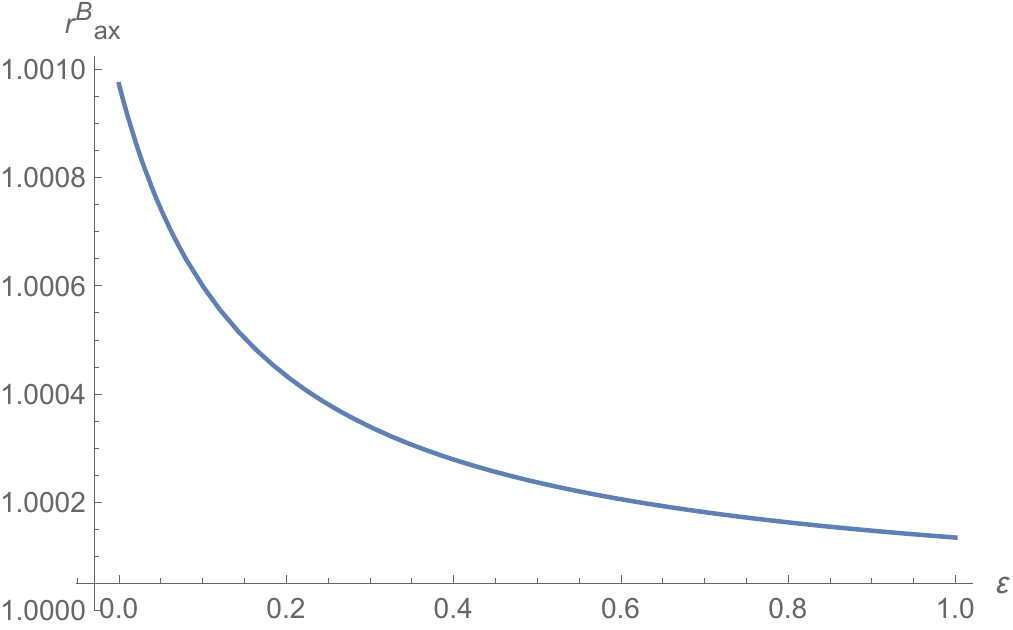}
%\vspace*{1cm}
\includegraphics[width=0.47\linewidth]{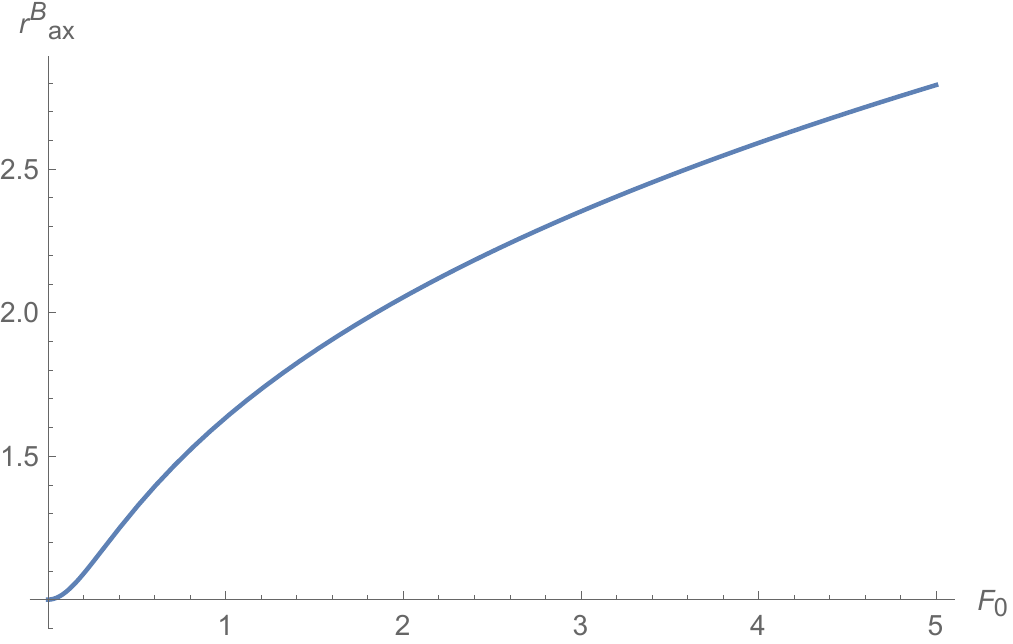}
\end{center}
\caption{Variation of the axial position of the isomagnetic axis, $r^B_{ax}$,  of the Solov'ev equilibrium of Fig. \ref{SolF} through the parameters $\delta$, $\epsilon$ and $F_0$ related, respectively, to the elongation, aspect ratio, and vacuum toroidal magnetic field.} 
                                                          \label{isoF}
\end{figure}

\begin{figure}[h]
%\vspace{-0.4cm}
\begin{center}
\includegraphics[width=0.49\linewidth]{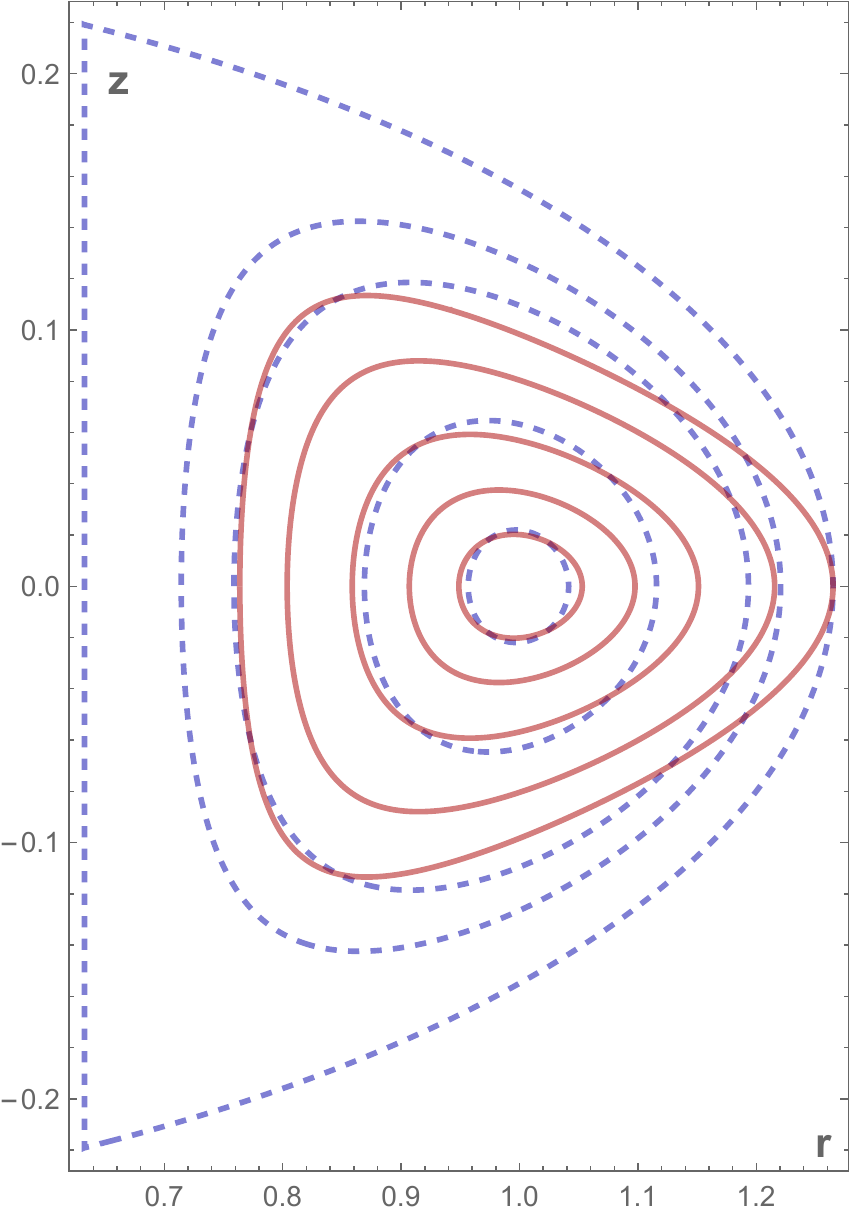}
\end{center}
\caption{Isomagnetic contours on the poloidal plane (red-continuous curves) for the equilibrium of Fig. \ref{SolF}, with the outer curve touching the separatrix. The blue-dashed curves represent the respective intersections of the magnetic surfaces with the poloidal plane.} 
                                                          \label{isoin}
\end{figure}

The equilibrium has nested isomagnetic surfaces without forming  a separatrix with the isomagnetic axis located at ($z=0$, $r=r^B_{ax}$);  the radial coordinate, $r^B_{ax}$,  can be obtained analytically even in the absence of symmetry in terms of the free parameters  $\delta$, $\epsilon$ and $F_0$. As shown in Fig.  \ref{isoF}, $r^B_{ax}$ strongly increases with $F_0$ and weakly decreases with $\delta$ and $\epsilon$.
Therefore, pending on the values of the free parameters, the position of the isomagnetic axis can be either inside or outside the magnetic separatrix. An example of the former case associated with the equilibrium of Fig. \ref{SolF} is given in Fig. \ref{isoin}; in the latter case another example of a 3D equilibrium is given in Fig. \ref{ex1iso}.

\section{The setting of 3D equilibria}
 We now consider 3D equilibria with anisotropic pressure governed by the equations:
\beq
\label{3Deq} 
 \nabla\cdot\bPc +  \bB\times \bj=0\,, \quad \nabla \times \bB = \bj \,, \quad \nabla\cdot\bB=0\,,
\eeq
where $\bPc=P_\perp \bIc + (P_\paral-P_\perp)/B^2 \bB \bB$ is the  pressure tensor consisting of
one element representing the parallel to the magnetic field pressure component  ($P_\paral$) and two equal perpendicular components ($P_\perp$), with $\bIc$ the identity tensor. Here, dimensionless quantities are employed too with the current density normalized by $B_a/(\mu_0 r_a)$. In \cite{LoSp}, a special class of solutions was identified, namely:
\beq
\label{sol}
P_\perp=P_0 -\frac{1}{2} B^2\,, \quad P_\paral=P_0 + \frac{1}{2} B^2\,,
\eeq
where $P_0$ is a constant pressure,  valid for any magnetic field. In view of (\ref{sol}) any divergence-free magnetic field, satisfies the force-balance equation. However, as stated in [4], this MHD solution does not represent confined plasmas, since $P_0$ is a constant pressure. In the above context, we introduce the following magnetic-field representation:
\beq
\label{Brep}
\bB=\nabla(\phi+ w(\phi))\times \nabla U(r,\phi,z) + I(r,z)\nabla \phi\,,
\eeq 
where the functions $U(r,f,z)$, $w(\phi)$ and $I(r,z)$ are arbitrary. This representation is advantageous in satisfying the third of Eqs. (\ref{3Deq}) identically.
The free function $U$ in (\ref{Brep}) is specified by toroidally perturbing the Solove'v solution (\ref{Sol}) as 
\beq
\label{Solp}
U(r,\phi,z)=z^2[r^2-\epsilon(1+g(\phi))]+\frac{\delta^2(1+h(\phi))}{4}(r^2-1)^2. 
\eeq
Since the form of (\ref{Solp}) is based on (\ref{Sol}), it may be expected that the perturbed equilibrium will retain certain   organized structure of the respective axisymmetric Solov'ev configuration.    
If the function $U$ is required to satisfy the equation $\nabla U\cdot \bB=0$ it turns out that the equilibrium should be axisymmetric; therefore $U$ can not label potential magnetic surfaces of 3D equilibria. 
In addition,  we make the following choices  for the periodic functions $g(\phi)$, $h(\phi)$ and $w(\phi)$ and the axisymmetric function $I(r,z)$:
\beq
\label{choice}
s(\phi)=c_{s} \cos (m_{s} \phi) + d_s \sin(n_s \phi)  \quad (s=g,h,w)\,, \quad I(r,z)= (F_0^2+ 4 \epsilon \psi_{ax})^{1/2},
\eeq 
 with $\psi_{ax}$ given by (\ref{Sol}). There are twelve free real parameters ($c_s$, $m_s$, $d_s$ $n_s\,:\  s=g,h,w$), in addition to  $\delta$, $\epsilon$ and $F_0$.
 % the values of $m_s$ and $n_s$ can be restricted to non-negative  without loss of generality.
 
 To examine whether the equilibria determined by Eqs. (\ref{sol})-(\ref{choice}) can have closed and nested magnetic surfaces we have traced out the magnetic field lines on the basis of the equation $d\bx(l)/dl=\bB(\bx)$, where $l$ is the arc-length associated with the vector $\bx$ tangential to $\bB$. The respective scalar equations in cylindrical coordinates are $dr/dl=B_r$, $d\phi/dl=B_\phi/r$ and $dz/dl=B_z$. Considering $r$ and $z$ as functions of $\phi$ we take the couple of ODEs: $dr/d\phi=r(\phi)B_r(r(\phi),\phi,z(\phi))/B_\phi(r(\phi),\phi,z(\phi))$ and $dz/d\phi=r(\phi)B_z(r(\phi),\phi,z(\phi))/B_\phi(r(\phi),\phi,z(\phi))$; they have been solved numerically making 400 toroidal revolutions with initial conditions $r(0)=r_0$ and 
$z(0)=z_0$ with ($r_0, z_0$) the coordinates of several points inside the plasma region. In a similar way can be traced out  a current-density line. Also, we have made a pertinent poloidal-cross-section consideration which is presented in  Appendix A.

\section{Construction of 3D equilibria}
\begin{figure}[h]
%\vspace{-0.4cm}
\begin{center}
\includegraphics[width=0.47\linewidth]{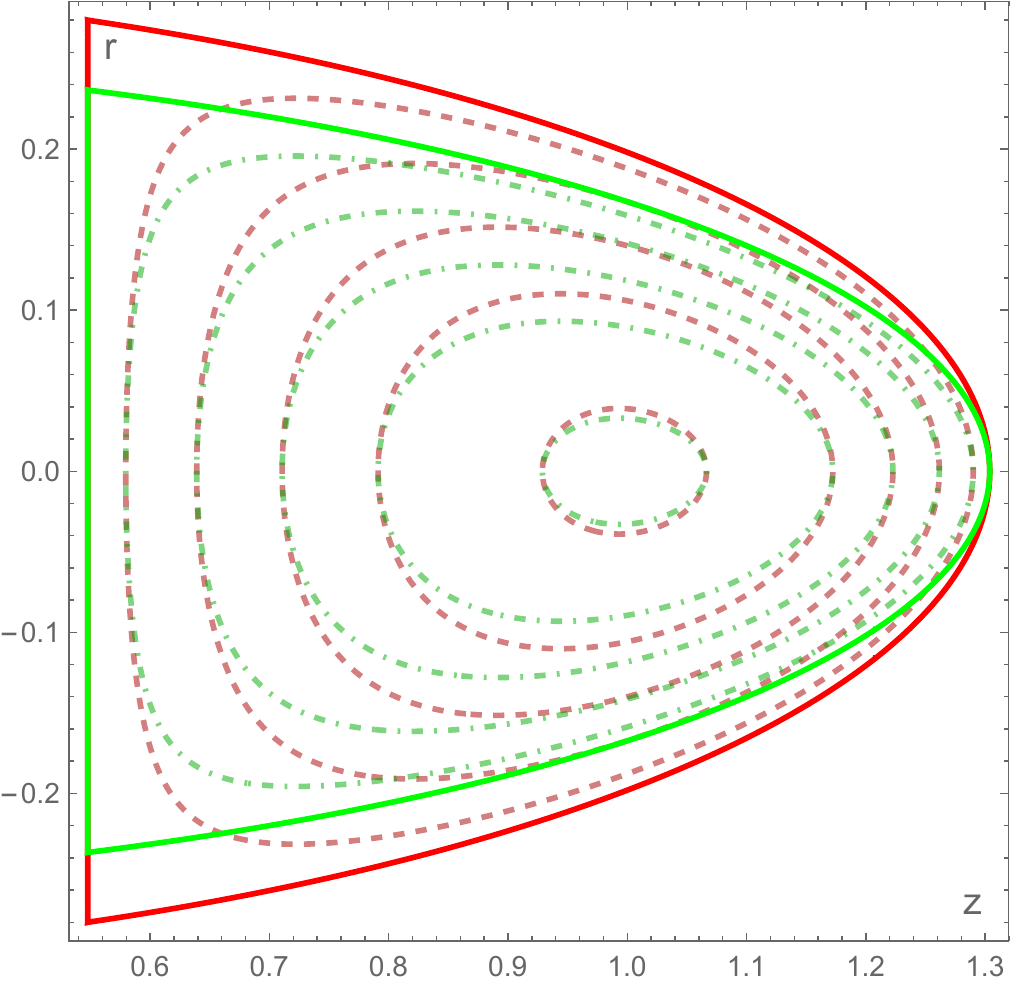}
\end{center}
\caption{Intersections of the magnetic surfaces (red, dashed contours) and of the current-density surfaces (green, dot-dashed contours) with the poloidal plane $\phi=0$ for an equilibrium with $\delta=0.4$, $\epsilon=0.3$, $F_0=3.5$, $c_s=d_s=0\ (s=g,w)$, $c_h=0.4$, $m_h=5$ and $d_h=0$. The  outer continuous contours correspond to  the respective separatrices. } 
                                                          \label{j_B_poloidal}
\end{figure}

 Applying the above procedure  we have constructed several equilibria wich showcase  closed and nested toroidal magnetic surfaces. They all have a magnetic axis located, as in the  axisymmetric case,  at ($r=1,\ z=0$) independently of $\phi$;  also, they form  a separatrix similar in shape with the axisymmetric one,  which, depending on the values of the free parameters and $\phi$, can be either inside or outside  the axisymmetric magnetic separatrix (cf. Appendix A).  An example on the poloidal cross-section $\phi=0$ is shown in Fig. \ref{j_B_poloidal}. 
The current-density surfaces, not coinciding with the magnetic surfaces, can be closed and nested too. They form a separatrix which coincides with the axisymmetric magnetic separatrix as demonstrated in the example of Fig. \ref{j_B_poloidal}. The equilibria exhibit  closed and nested isomagnetic surfaces. The radial position, $r^B_{ax}$, of the isomagnetic axis depending on the free parameters $\delta$, $\epsilon$ and $F_0$ in a similar way as in the axisymmetric case (Fig. \ref{isoF}), it also depends on the perturbation parameters $c_h$ and $d_h$
and it can  vary with the toroidal coordinate $\phi$. 

\begin{figure}[h]
%\vspace{-0.4cm}
\begin{center}
\includegraphics[width=0.35\linewidth]{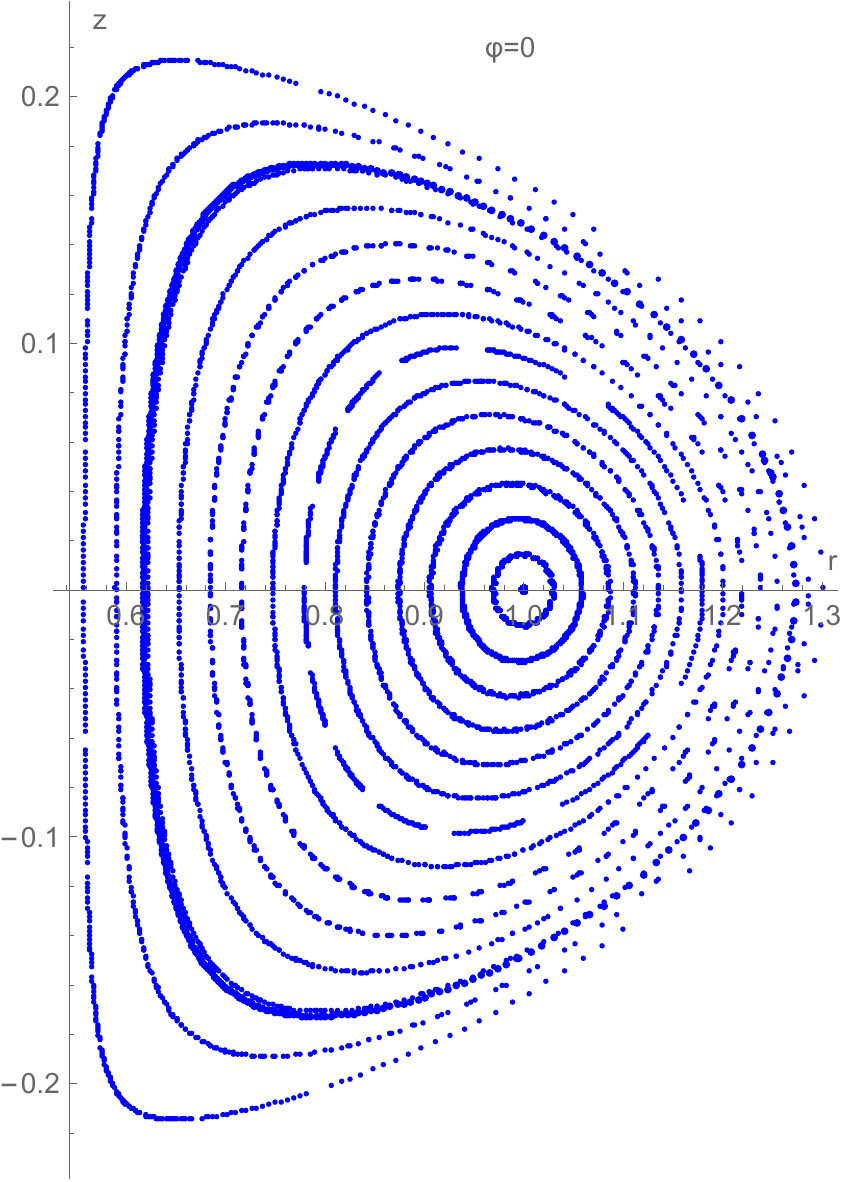}
\hspace{0.3cm}
\includegraphics[width=0.60\linewidth]{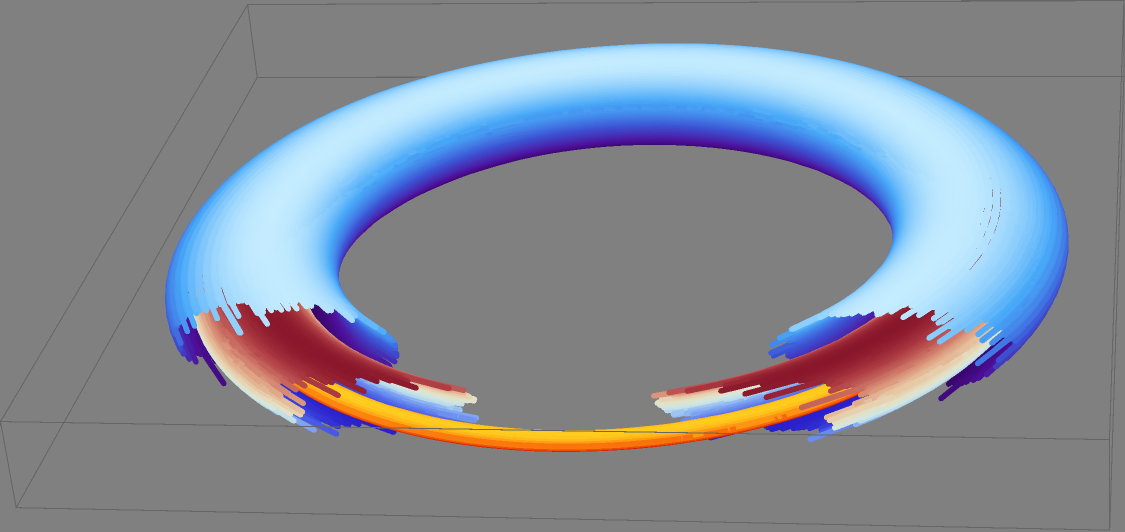}
\end{center}
\caption{Left panel: Poincar\'e plot of magnetic-field lines  on the poloidal plane $\phi=0$. Right panel: Respective 3D magnetic surfaces. } 
                                                          \label{ex1B}
\end{figure}

\begin{figure}[h]
%\vspace{-0.4cm}
\begin{center}
\includegraphics[width=0.35\linewidth]{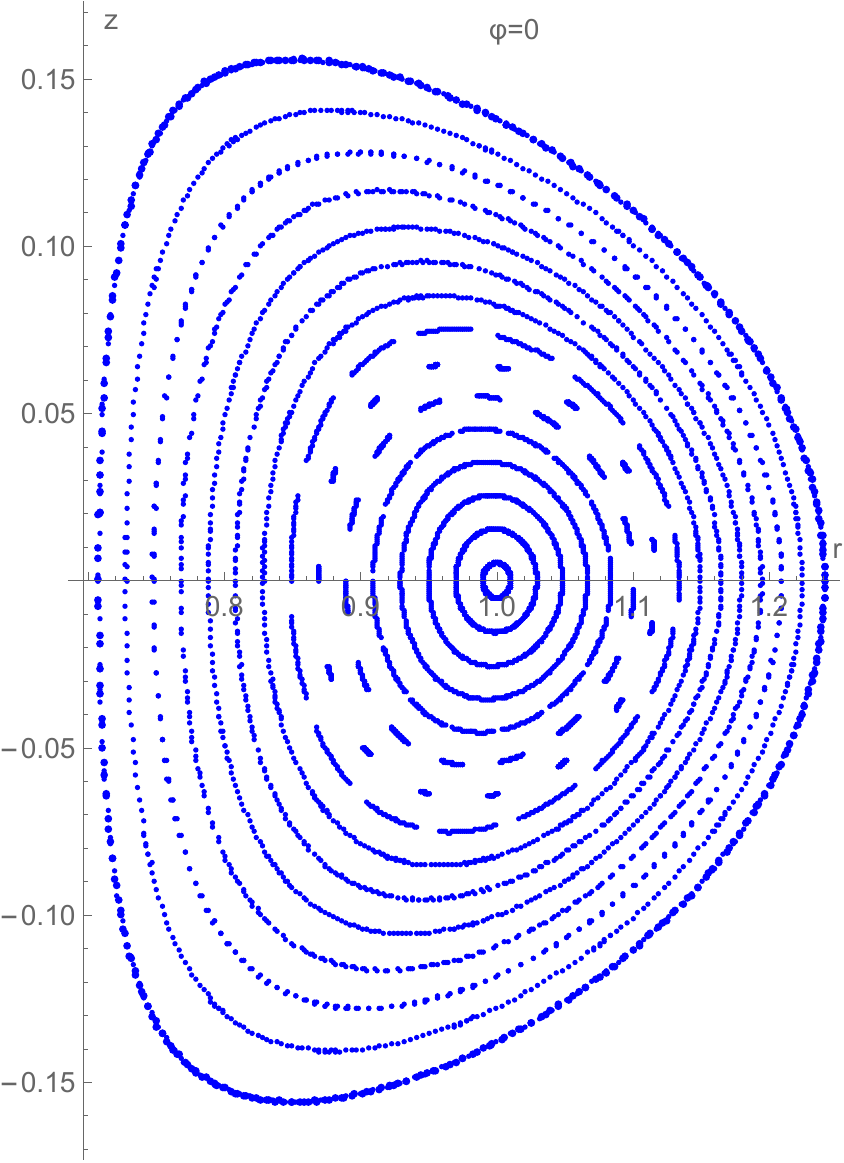}
\hspace{0.3cm}
\includegraphics[width=0.60\linewidth]{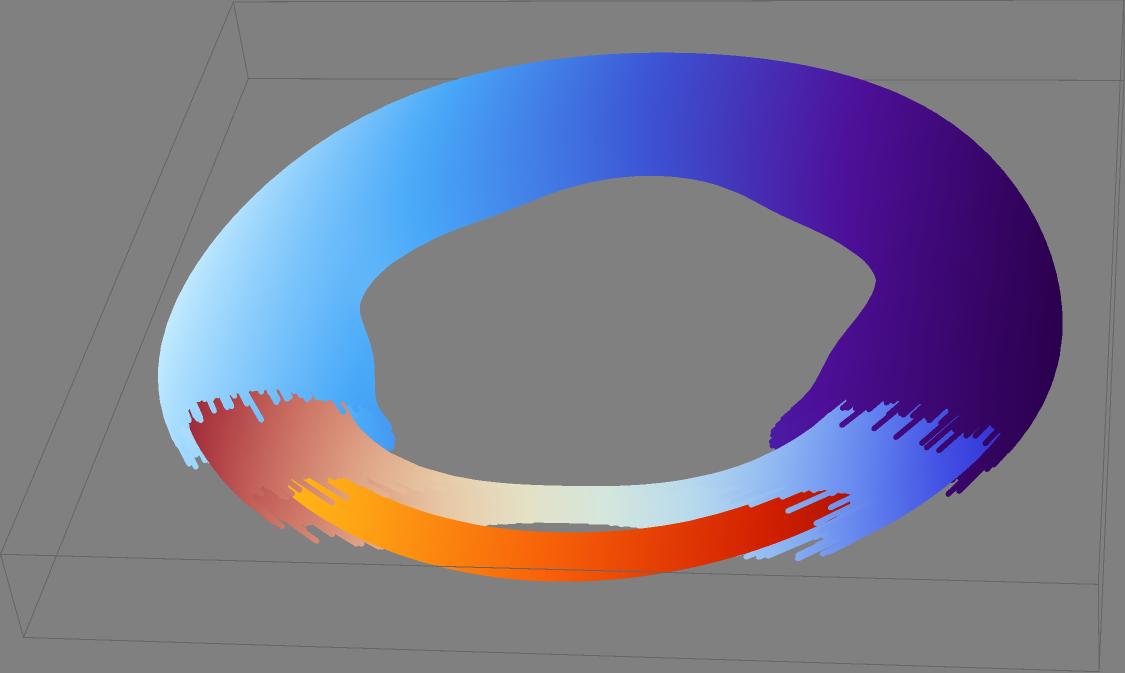}
\end{center}
\caption{Left panel: Poincar\'e plot for the current-density lines of the equilibrium of Fig. \ref{ex1B} on the poloidal plane $\phi=0$. Right panel: Respective 3D current-density surfaces.}
%an equilibrium with $\delta=0.4$, $\epsilon=0.3$, $F_0=3.5$, $c_s=d_s=0\ (s=g,w)$, $c_h=0.4$, $m_h=5$ and $d_h=0$. 
                                                          \label{ex1j}
\end{figure}

\begin{figure}[h]
%\vspace{-0.4cm}
\begin{center}
\includegraphics[width=0.50\linewidth]{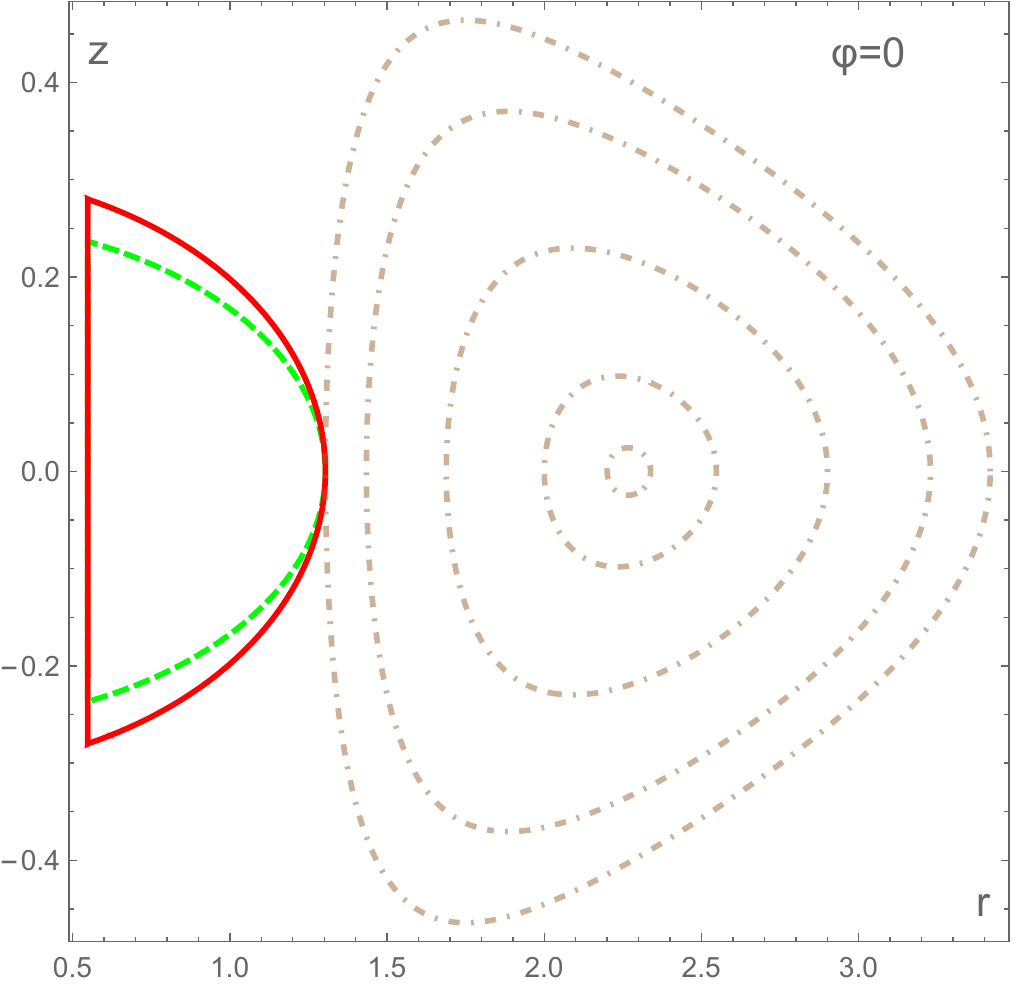}
\end{center}
\caption{Isomagnetic contours (brown, dot-dashed) on the poloidal plane $\phi=0$ for the equilibrium of Figs. \ref{ex1B} and \ref{ex1j} with the isomagnetic axis located outside the magnetic separatrix  indicated by the red-continuous curve.  The green-dashed curve indicating the cut of a current-density surface with the poloidal plane   coincides with the respective curve of the axisymmetric separatrix.}  
                                                          \label{ex1iso}
\end{figure}

\begin{figure}[h]
%\vspace{-0.4cm}
\begin{center}
\includegraphics[width=0.45\linewidth]{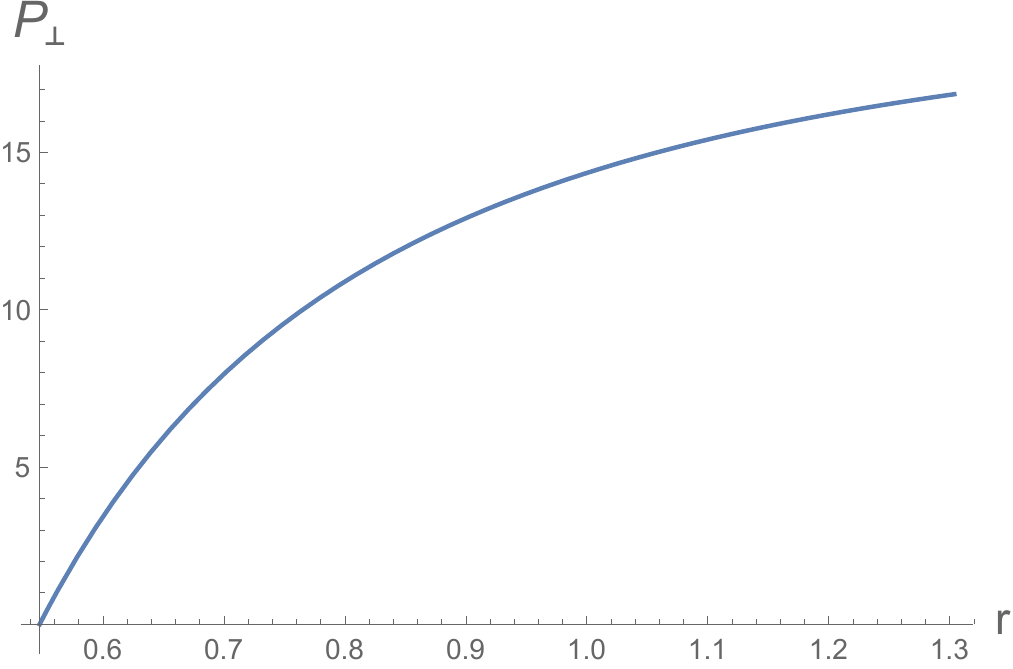}
\includegraphics[width=0.45\linewidth]{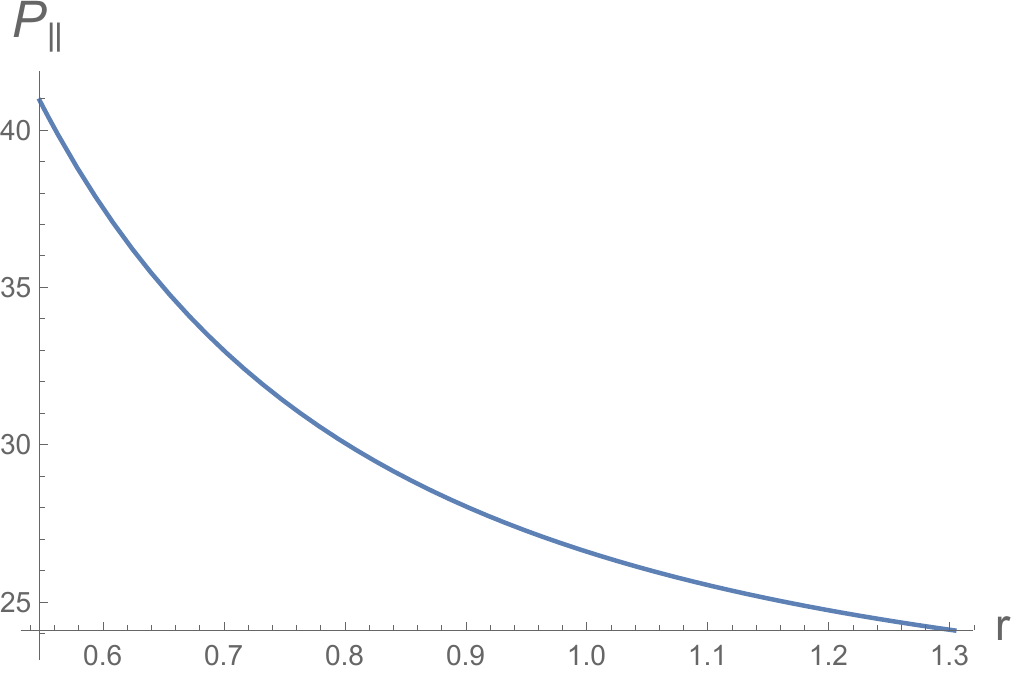}
\includegraphics[width=0.45\linewidth]{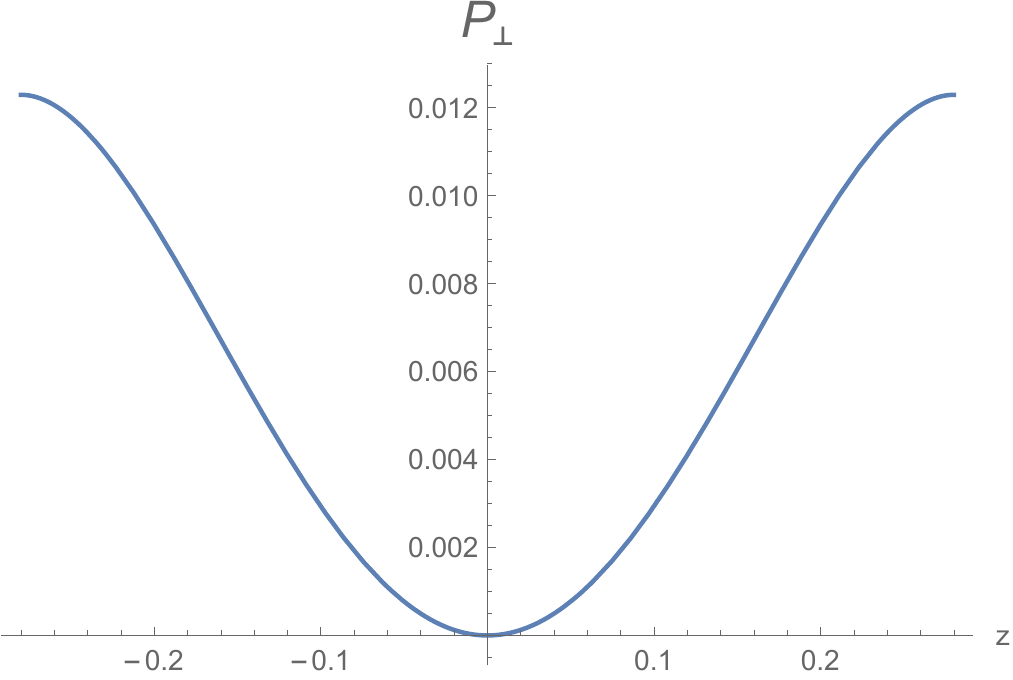}
\includegraphics[width=0.45\linewidth]{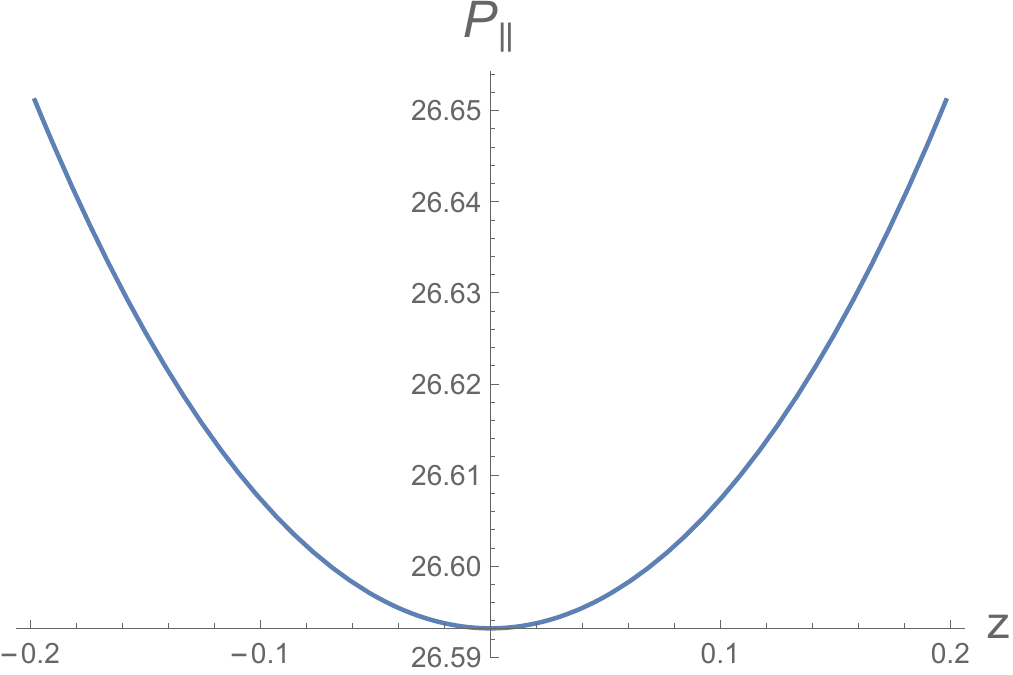}
\end{center}
\caption{Upper panel: Radial variation of the pressure elements, $P_\perp$ and $P_\paral$, from the inner to the outer point of the magnetic separatrix on the middle-plane $z=0$ for the equilibrium of Figs. \ref{ex1B} and \ref{ex1j}. Lower panel: Respective $z$-profiles at the radial position, $r=1$, of the magnetic axis.}  
                                                          \label{P_profiles}
\end{figure}

A Poincar\'e plot  for $\bB$ at $\phi=0$ together with the 3D magnetic surfaces for the  equilibrium of Fig. \ref{j_B_poloidal} are  presented  in Fig. \ref{ex1B}  . 
Respective plots for the current-density lines are given in Fig. \ref{ex1j}. 
Owing to the small value 0.4 of the perturbation parameter $c_h$ the magnetic surfaces look like axisymmetric. However, visible toroidal deformation appears in the current-density surfaces. Since the vacuum toroidal magnetic field is strong ($F_0=3.5$) the isomagnetic axis is located outside the magnetic separatrix (Fig. \ref{ex1iso}).
 This implies that the existence of closed and nested isomagnetic surfaces inside the plasma region is not necessary for the existence of respective closed and nested magnetic surfaces. The isobaric surfaces of $P_\perp$ and $P_\paral$ are identical with the isomagnetic surfaces (cf. (\ref{sol})). Radial profiles on the middle-plane $z=0$ and perpendicular ones at the radial position of the magnetic axis, $r=1$, are given  in Fig. \ref{P_profiles}. The value of $P_0$ has been assigned by requiring that the value of $P_\perp$ at the inner radial point of the separatrix on the middle-plane $z=0$ vanishes. This choice guarantees that $P_\perp$ and $P_\paral$ are non-negative throughout the whole plasma region. As can be seen in Fig. \ref{P_profiles} the $z$-variation of $P_\perp$ and $P_\paral$ is very small. Also, the  toroidal variation of $P_\perp$ and $P_\paral$ is negligible. 

\begin{figure}[h]
%\vspace{-0.4cm}
\begin{center}
\includegraphics[width=0.30\linewidth]{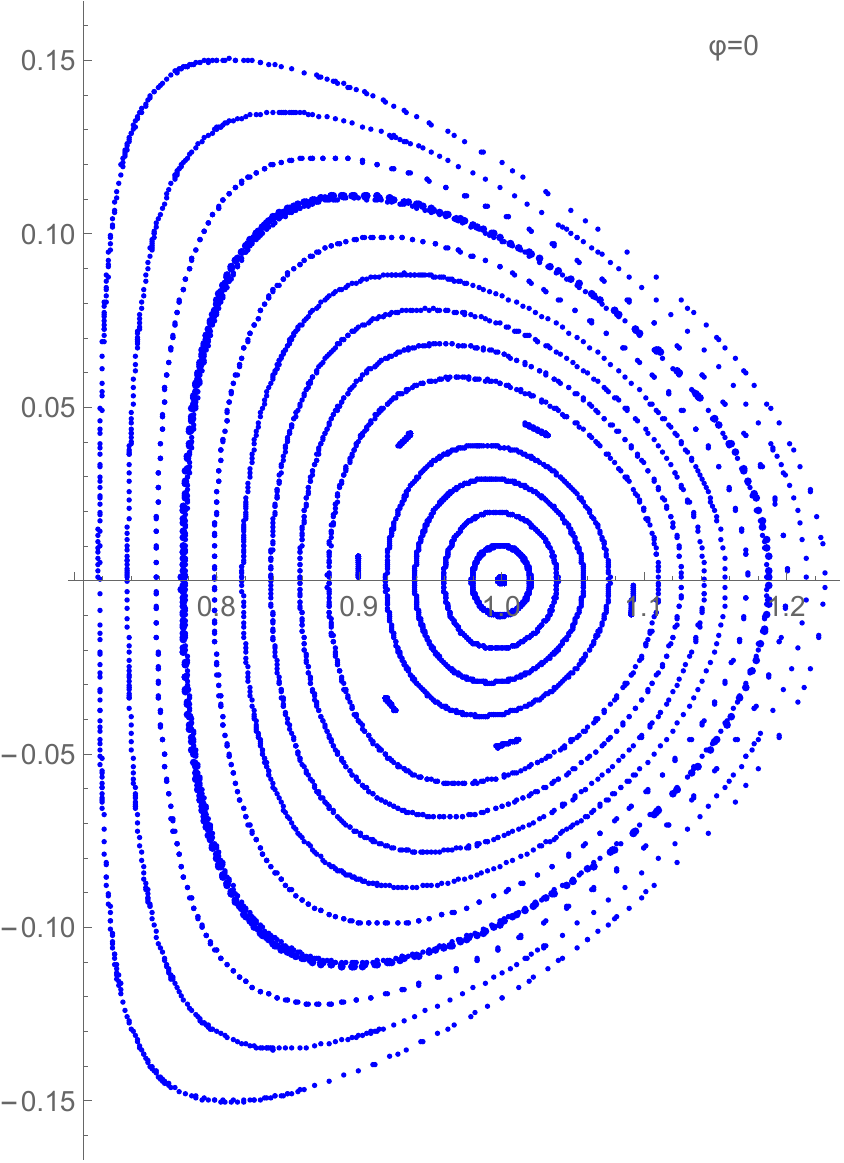}
\includegraphics[width=0.30\linewidth]{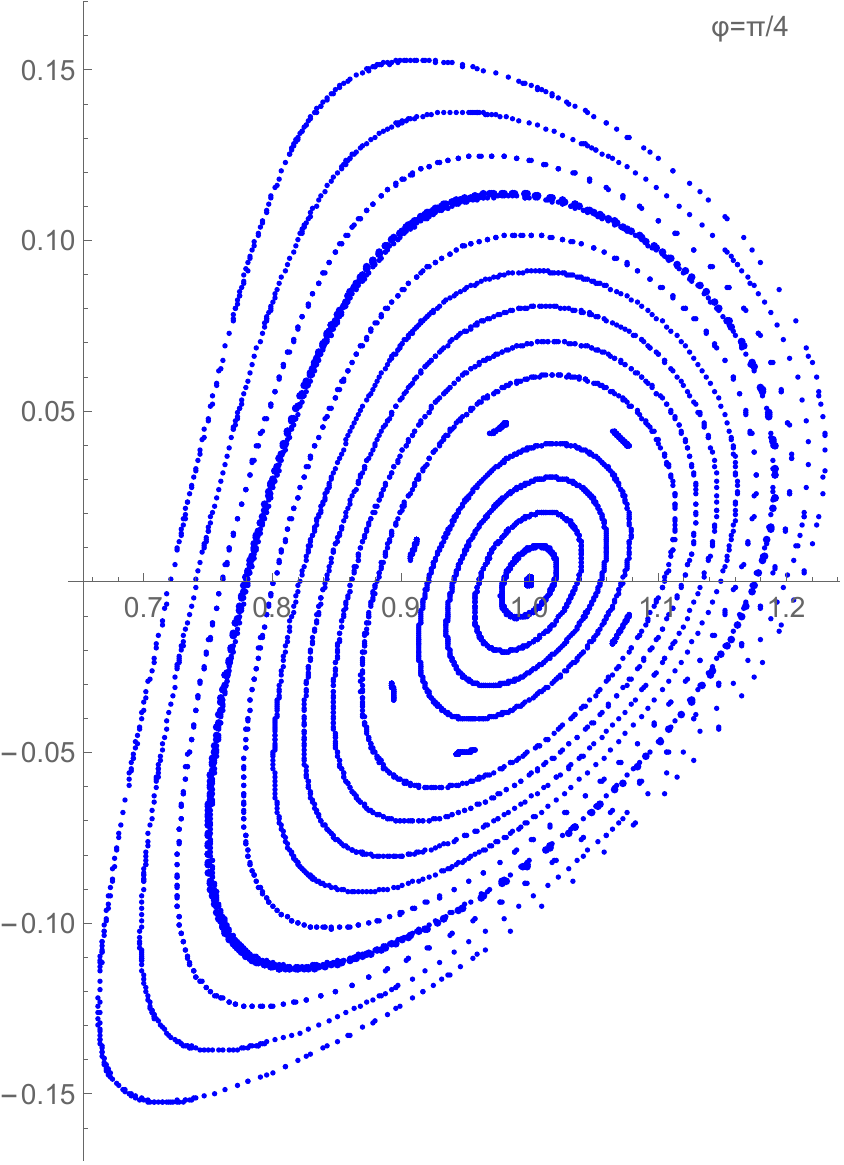}
\includegraphics[width=0.30\linewidth]{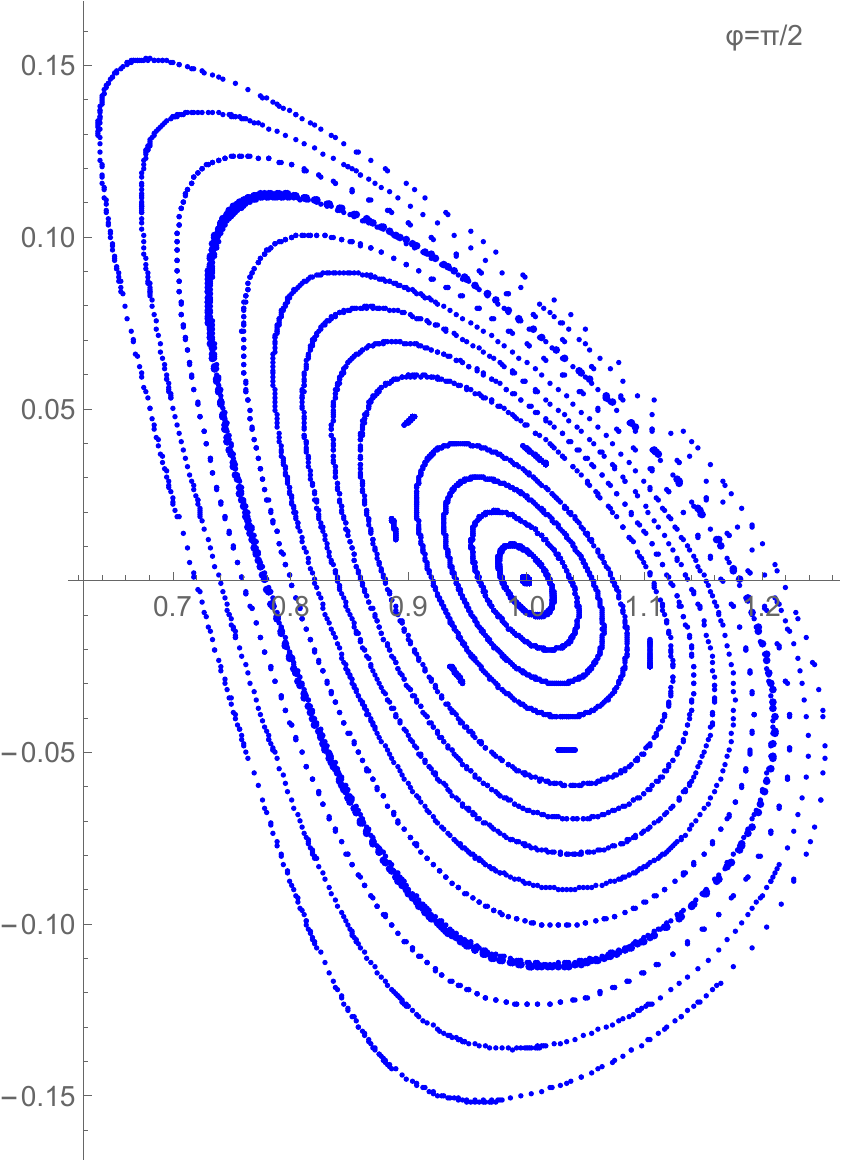}

\vspace{5mm}
\includegraphics[width=0.90\linewidth]{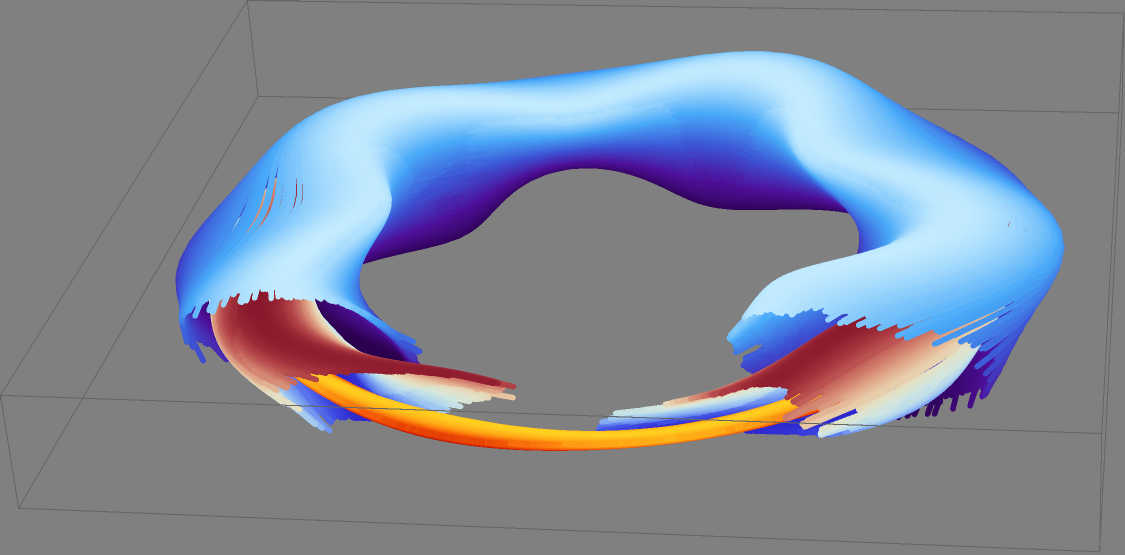}
\end{center}
\caption{Poincar\'e plots of a strongly toroidally asymmetric equilibrium on three poloidal cross-sections and the respective 3D depiction of the magnetic surfaces for $\delta=0.4$, $\epsilon=0.5$, $F_0=3.5$, $c_s=d_s=0\ (s=h,w)$, $c_g=20$, $m_g=5$ and $d_g=0$.}  
                                                          \label{ex2}
\end{figure}

\begin{figure}[h]
%\vspace{-0.4cm}
\begin{center}
\includegraphics[width=0.45\linewidth]{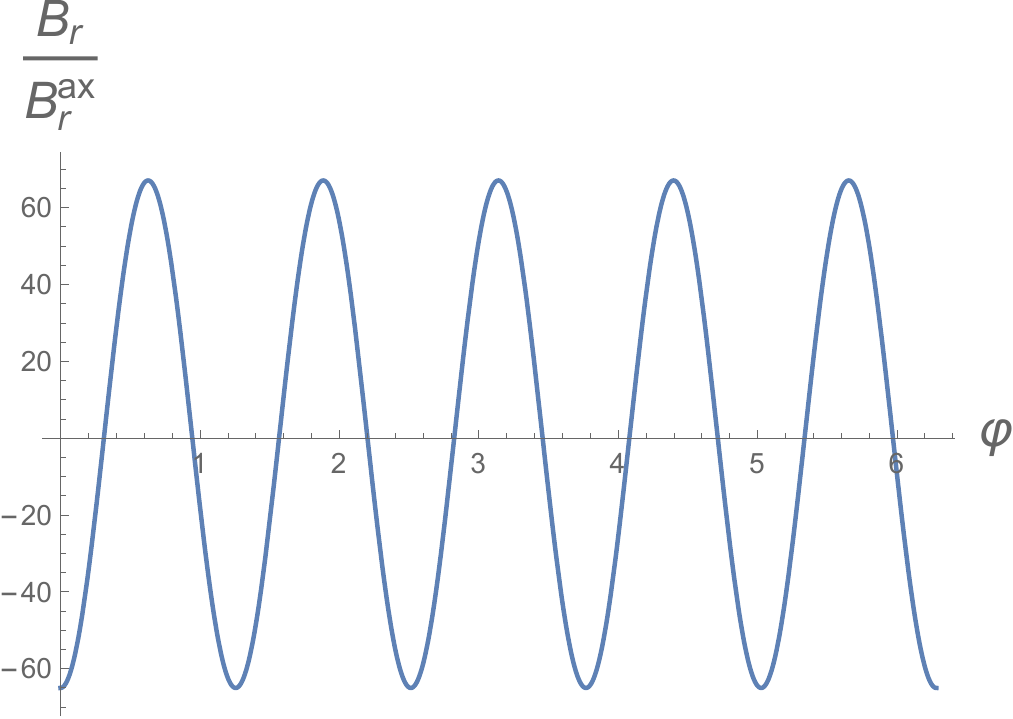}
\includegraphics[width=0.45\linewidth]{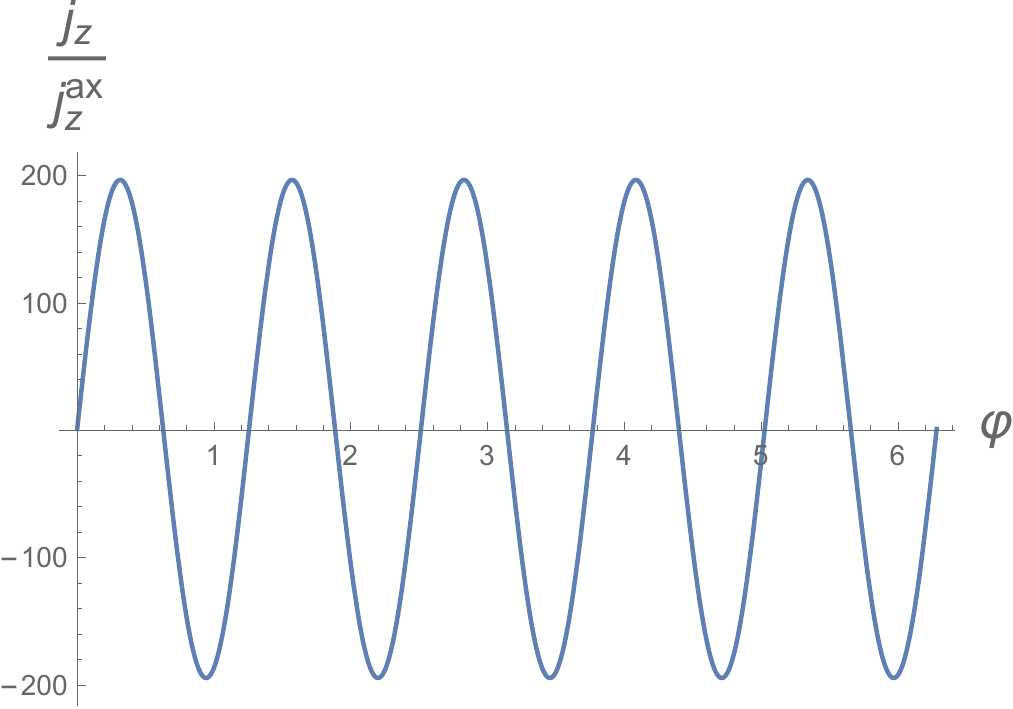}
\end{center}
\caption{Toroidal variation of the $r$-component of the magnetic field and of the $z$-component of the current density normalized by the respective axisymmetric quantities at the point ($r=0.807,\ z=0.2$).}  
                                                          \label{ex2_B_j}
\end{figure}

Another equilibrium strongly perturbed via the function $g(\phi)$ is shown in Fig. \ref{ex2}.
Quantitatively, the maxima of the components of $\bB$ is one order of magnitude   and those of  $\bj$ two orders of magnitude larger than those in the respective axisymmetric Solov'ev equilibrium (Fig. \ref{ex2_B_j}). 
The isomagnetic axis of this equilibrium is also located outside the magnetic separatrix. 

\begin{figure}[h]
%\vspace{-0.4cm}
\begin{center}
\includegraphics[width=0.45\linewidth]{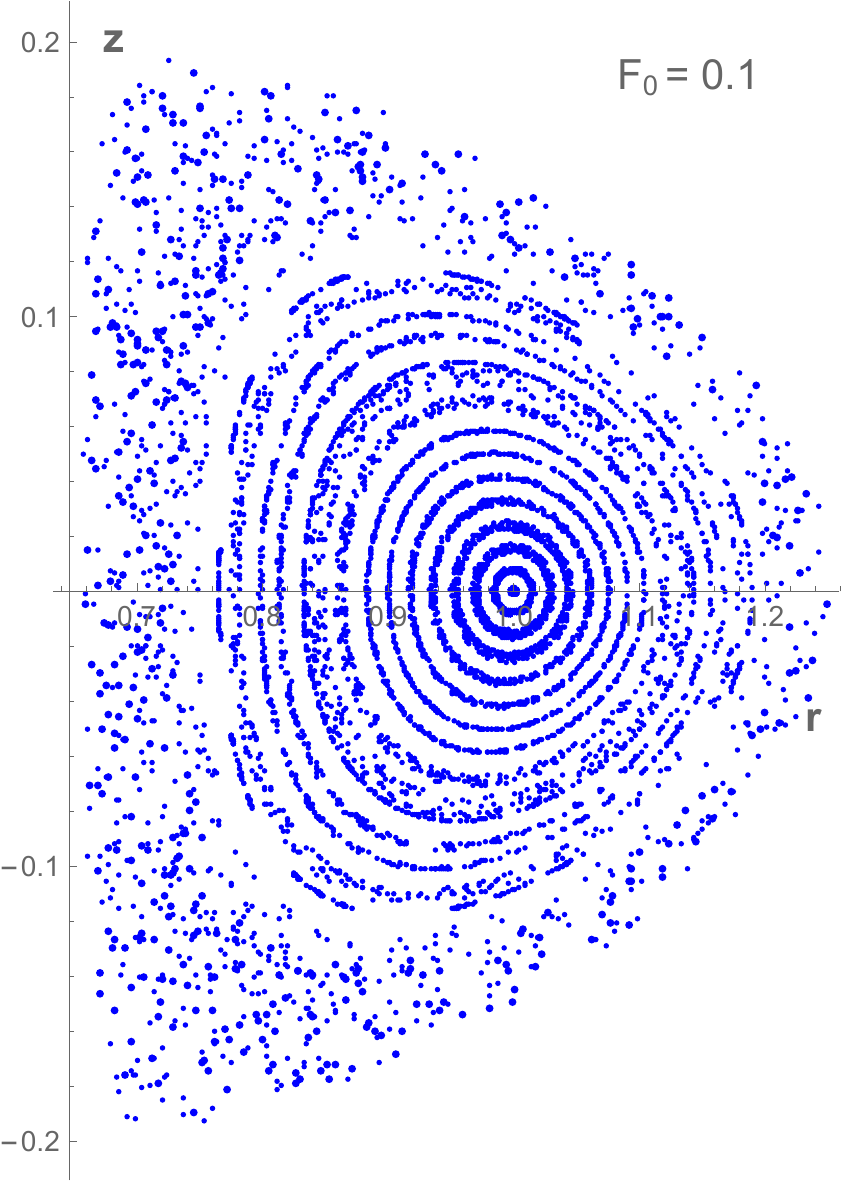}
\includegraphics[width=0.45\linewidth]{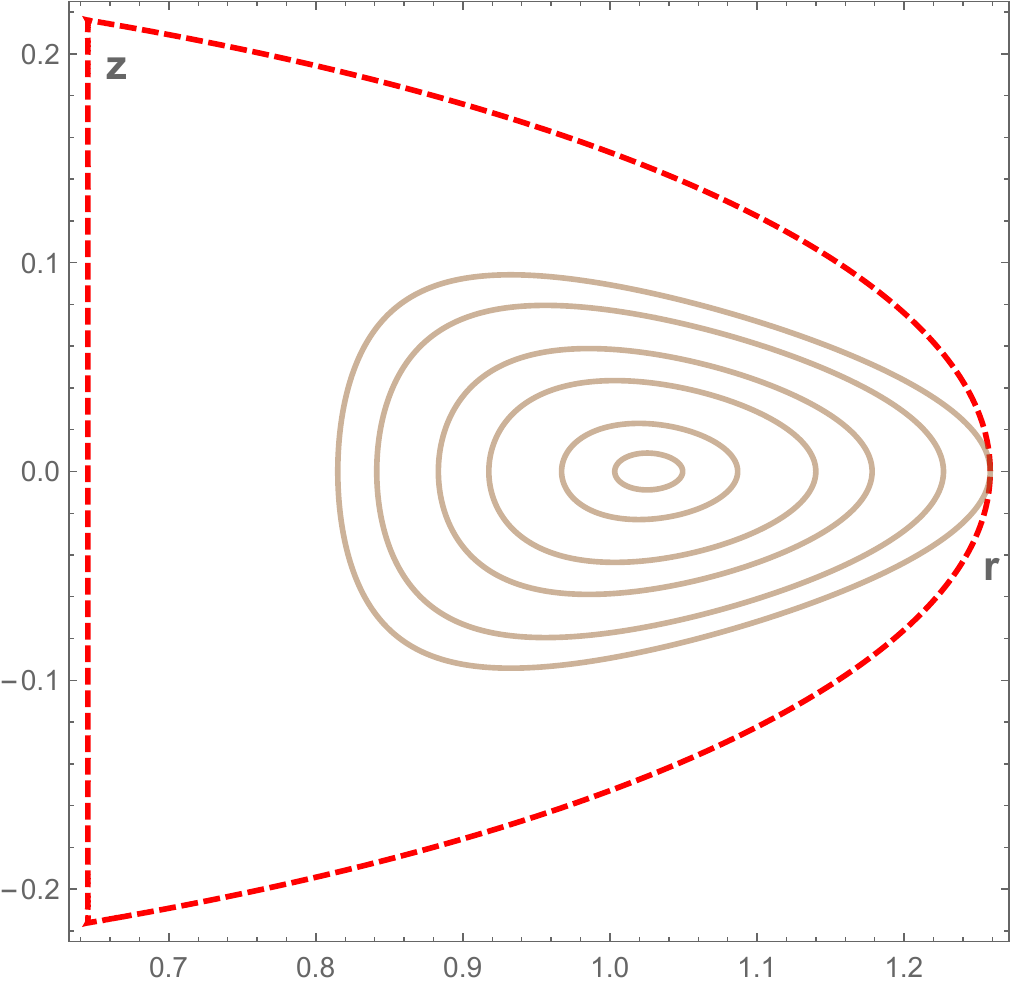}
\includegraphics[width=0.60\linewidth]{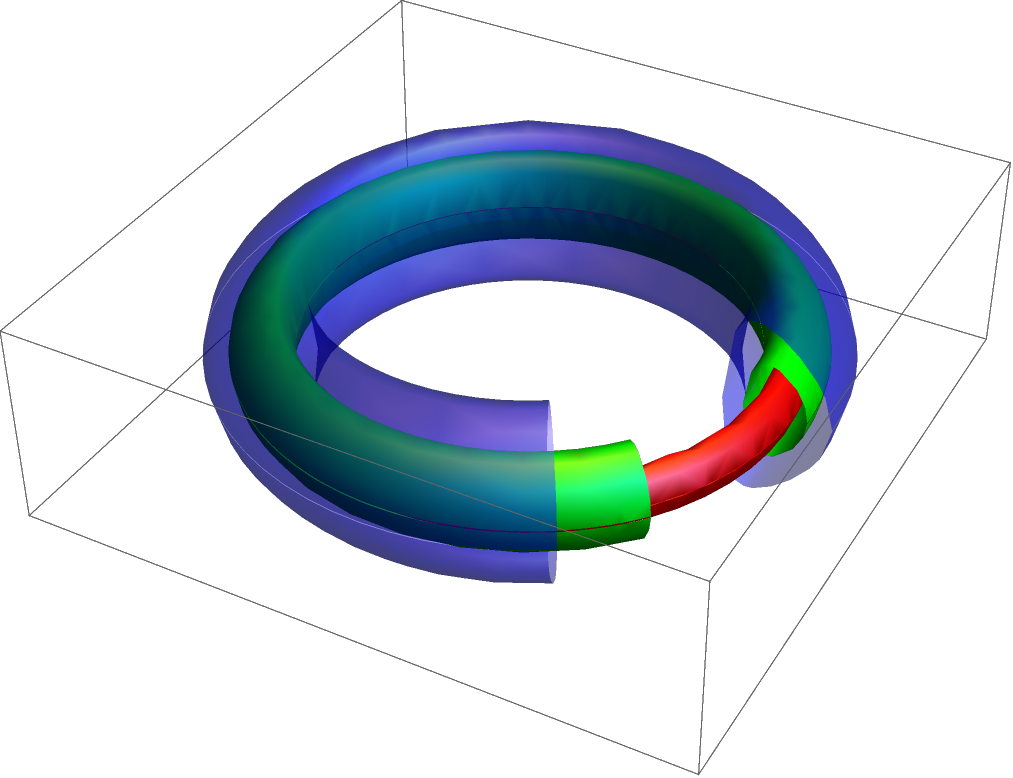}
\end{center}
\caption{Upper-left panel: Poincar\'e  plot of the magnetic-field lines for $\phi=0$ of an equilibrium having a large stochastic outer subregion, while in the central region the magnetic surfaces remain closed and nested for $\delta=\epsilon=0.4$, $F_0=0.1$, $c_g=0.04$, $m_g=1$, $d_g=0$ and  $h(\phi)= w(\phi) \equiv 0$ .
Upper-right panel: The respective closed and nested isomagnetic curves (brown-continuous) with the isomagnetic axis positioned inside the separatrix indicated with the red-dashed curve. Lower-panel: 3D depiction of the isomagnetic surfaces.}  
                                                          \label{ex3}
\end{figure}

\begin{figure}[h]
%\vspace{-0.4cm}
\begin{center}
\includegraphics[width=0.45\linewidth]{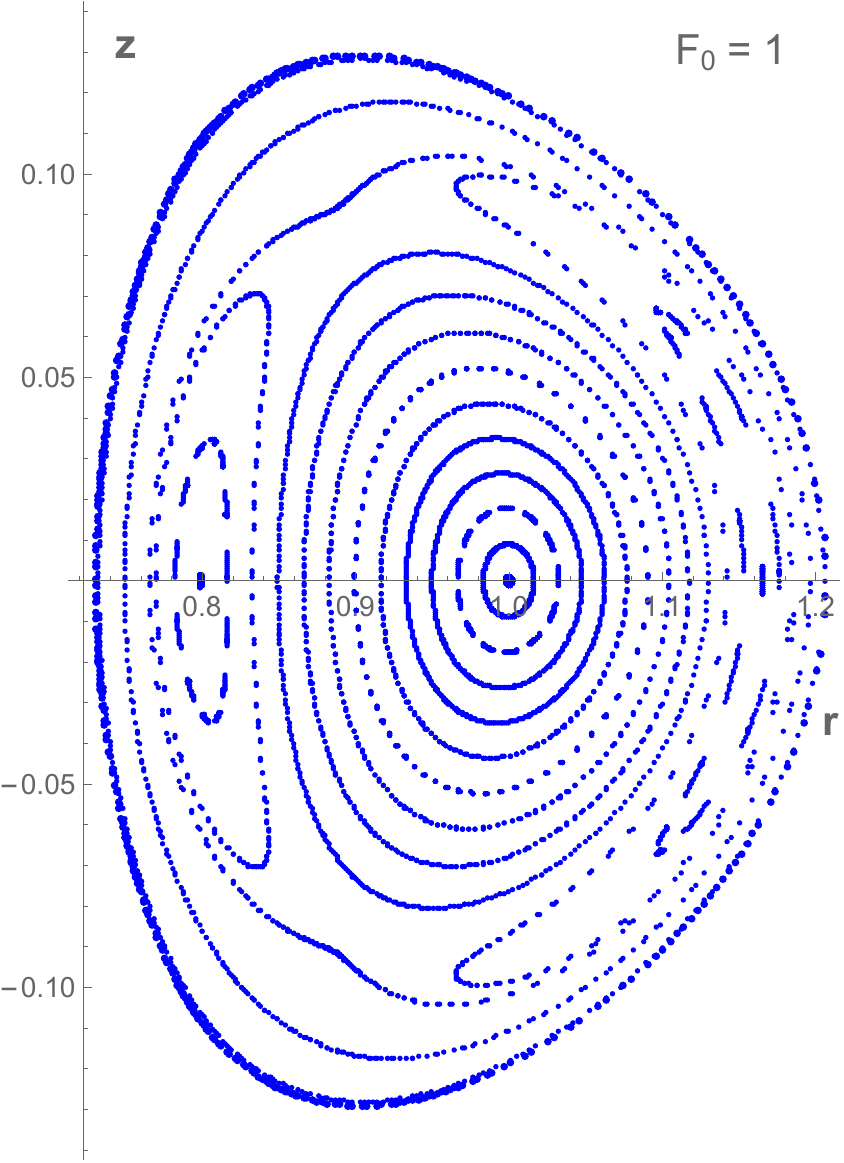}
\includegraphics[width=0.45\linewidth]{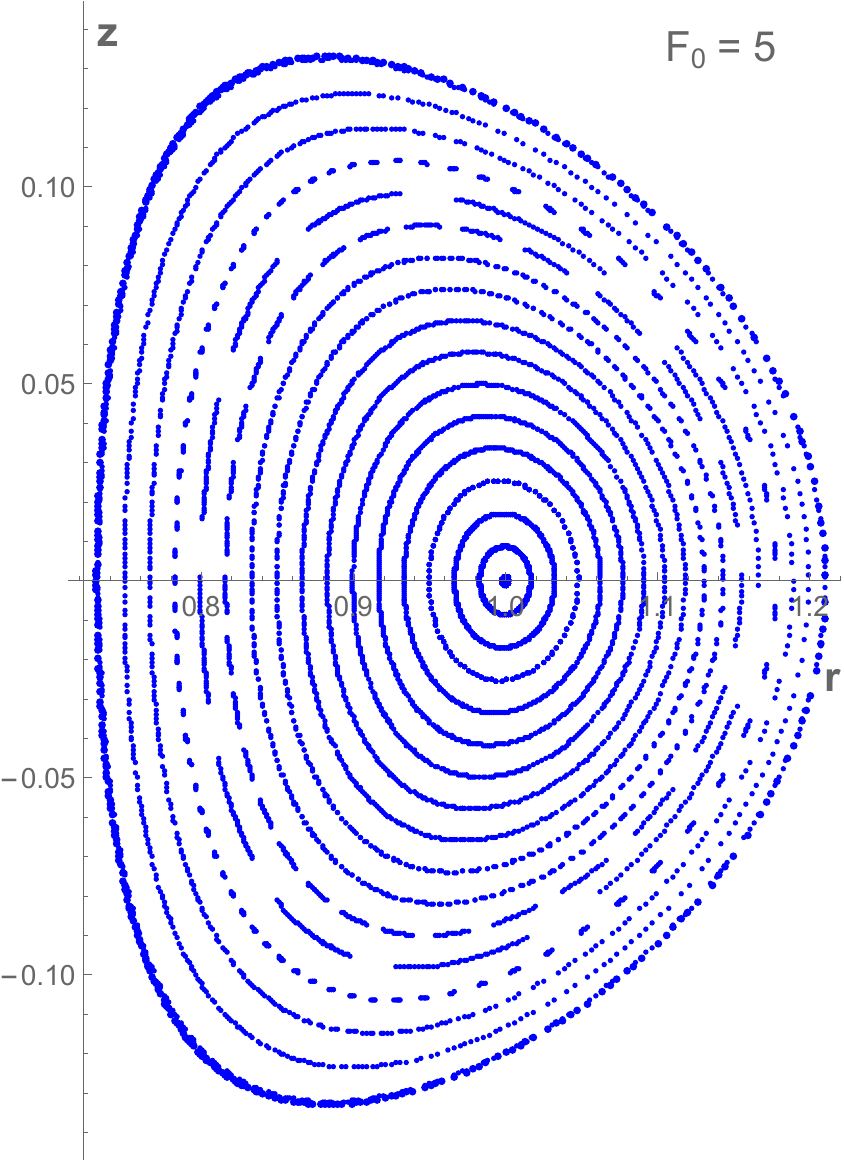}
\end{center}
\caption{Poincar\'e  plots of the magnetic-field for $\phi=0$ of two equilibria having identical parametric values with those of Fig. \ref{ex3} except $F_0$ which takes larger values. As $F_0$ increases the stochastic region in the configuration of the upper-left panel of Fig. \ref{ex3} shrinks and magnetic islands are formed ($F_0=1$).  Eventually, for sufficiently large values of $F_0$ ($F_0 =5$ here), the magnetic islands disappear.  }    
                                                          \label{ex3_F0}
\end{figure}

For certain parametric values a  subregion appears in the outer part of the plasma region close to the separatrix with stochastic magnetic field lines or/and magnetic islands. An example is shown in the upper-left panel of Fig. \ref{ex3}.  The respective isomagnetic axis lies inside the separatrix as shown in the upper-right panel of Fig. \ref{ex3}  
together with isomagnetic contours.  3D  closed and nested isomagnetic surfaces   are depicted in the lower panel. The outer of the isomagnetic contours   in Fig. \ref{ex3} touches the separatrix. While in the region close to the separatrix the isomagnetic contours are closed and nested, the magnetic field in the same region is stochastic. Therefore, the existence of closed and nested isomagnetic surfaces within the plasma region is not sufficient for the existence of closed and nested magnetic surfaces.
The stochastic region enlarges when $\delta$, $|c_s|$, $|d_s|$ ($s=g,h,w$) and $m_w$, $n_w$  take larger values and shrinks when $\epsilon$, $m_s$, $n_s$ ($s=g,h$) and $F_0$ take larger values,  the latter parameter having the strongest impact (Fig. \ref{ex3_F0}).

\section{Conclusions}
 We have constructed a class of  3D toroidal equilibria with anisotropic pressure that showcase closed, nested magnetic surfaces via harmonic perturbations to the  axisymmetric Solov'ev equilibrium (Eqs. (\ref{sol}-\ref{choice})). The amplitude of the perturbations can be arbitrary and, therefore,  equilibria with strong toroidal asymmetry are possible (cf. lower panel of Fig. \ref{ex2}). The magnetic surfaces having a separatrix depart from the current-density surfaces which are closed and nested too and have a distinct separatrix   which  coincides with the axisymmetric magnetic one.  The equilibria exhibit closed and nested isomagnetic surfaces with  the isomagnetic axis, depending on the values of the free parameters,  positioned either inside or outside the magnetic separatrix. It has been demonstrated  that the existence of closed and nested isomagnetic surfaces inside the plasma region
%, related to quasisymmetry, 
is neither necessary nor sufficient for the existence of respective closed and nested magnetic surfaces. For certain values of the free parameters  an area of stochastic magnetic-field lines potentially including magnetic islands is  formed in the  outer region close to the separatrix, while closed and nested magnetic surfaces persist  in the central region. The extent of this area depends on the free parameters; in particular, it drastically shrinks as the vacuum toroidal magnetic field takes larger values. 

The set of Eqs. (\ref{sol}-\ref{choice}) can be employed to constructing 3D equilibria as perturbations of any axisymmetric solution of the Grad-Shafranov equation, e.g. the Herrnegger-Maschke solution \cite{He,Ma} having a  current-density profile peaked on the magnetic axis and vanishing on the boundary. Also, the potential  construction of  3D equilibria with closed,  nested toroidal magnetic surfaces and strong toroidal
asymmetry  for magnetically confined plasmas  remains an  open question which we will address in future studies.

\section*{Acknoweledgent}
This work was conducted in the framework of participation of the University of Ioannina in the National Programme for the Controlled
Thermonuclear Fusion, Hellenic Republic. 
%The authors would like to thank the anonymous reviewers for constructive comments, which have resulted in an improved version of the manuscript.
%The views and opinions expressed herein do not necessarily
%reflect those of the European Commission.

\section*{Appendix A: Poloidal cross-section consideration}
\renewcommand{\theequation}{A.\arabic{equation}}
\setcounter{equation}{0}
Here, we will make a 2D equilibrium consideration by replacing the toroidal coordinate $\phi$ by a parameter $a$  ranging in the interval $[0, 2\pi]$. Then, it is possible to find   a function $u(r,z;a)$ to the level sets of which  the magnetic field intersections with the poloidal plane $\phi=a$ are tangential, i.e. by solving analytically the equation $\nabla u\cdot \bB=0$. The solution for the magnetic field given by  (\ref{Brep}-\ref{choice})  (for $\phi=a$) is  
\beqa
u(r,z;a)&=&f(y)\,,  \nonumber \\
&   y=&4 r^2 z^2 - 2 r^2 \delta^2 + r^4 \delta^2 \nonumber \\ 
 & & - 4 z^2 \epsilon  \newline - 4 z^2 \epsilon g(a) - 
 2 r^2 \delta^2 h(a) + r^4 \delta^2 h(a)\,, 
\eeqa
where $f$ is an arbitrary smooth function of $y$.
An example of closed and nested $u$-lines on the poloidal plane $a=0$ are depicted in Fig. \ref{j_B_poloidal} (red-dashed lines). The outer continuous line  corresponds to a separatrix similar in shape with but lying outside the axisymmetric magnetic separatrix.
The respective level sets of a function $u_j(r,z)$ labelling the current-density intersections  with the plane $a=0$ can be found analytically too as solution of the equation $\nabla u_z\cdot \bj=0$. The closed and nested $u_j$ lines for the above example are also displayed in Fig. \ref{j_B_poloidal}. The outer of these lines corresponds to a current-separatrix which coincides with the axisymmetric magnetic one. 

It is clarified that the above 2D consideration   for various values of $a$, though helpful,  does not provide the actual equilibrium characteristics obtained by the 3D magnetic-field and current-density line tracing described in Sec. 3.


\begin{thebibliography}{99}

\bibitem{Su} C. Lo Surdo, \textit{Global magnetofluidostatic fields (an unsolved PDE problem)}, International Journal of Mathematics
and Mathematical Sciences {\bf 9}, 123 (1986).

\bibitem{SaYa} N. Sato and M. Yamada, \textit{Nested invariant tori foliating a vector field and its curl: Toward MHD equilibria and
steady Euler flows in toroidal domains without continuous Euclidean isometries}, J. Math. Phys. {\bf 64}, 081505 (2023)

\bibitem{Gr} H. Grad, \textit{Toroidal Containment of a Plasma}, Phys. Fluids {\bf 10}, 137 (1967). 
%\url{https://doi.org/10.1088/0741-3335/59/1/014012}.

\bibitem{St} T. H.  Stix, \textit{Magnetic Braiding in a Toroidal Plasma}, Phys. Rev. Let. {\bf 30}, 833 (1973).

\bibitem{Gr85} H. Grad, \textit{Theory and applications of the nonexistence of simple toroidal plasma equilibrium}, International Journal
of Fusion Energy {\bf 3}, 33 (1985).

\bibitem{Mo} H. K. Moffatt, \textit{Magnetostatic equilibria and analogous Euler flows of arbitrary complex topology. Part 1.
Fundamentals},  Journal of Fluid Mechanics {\bf 159}, 359  (1985).

\bibitem{BrLa} O. P. Bruno and P. Laurence, \textit{Existence of three-dimensional toroidal MHD equilibria with nonconstant pressure}, Communications on Pure and Applied Mathematics {\bf 49}, 717  (1996).

\bibitem{PaBa83} D. Palumbo, M. Balzano \textit{The unicity of isodynamic toroidal configurations}, Atti Accad. Sci. Lett.,  Ser. V, Parte I {\bf 4}, 127 (1984).

\bibitem{Pa86} D. Palumbo, \textit{Some properties of MHS equilibrium toroidal equilibria and non-existence of the isodynamic stellarator}, Atti Acc. Sc. Lett. 
{\bf IV} (parte I), 107 (1986).

\bibitem{Pa68} D. Palumbo, \textit{Some considerations on closed configurations of magnetohydrostatic equilibrium}, Nuovo Cimento B {\bf 53}, 507 (1968).

\bibitem{Hel}Per Helander, \textit{Theory of plasma confinement in nonaxisymmetric magnetic fields}, Rep. Prog. Phys. {\bf 77}, 087001 (2014).

\bibitem{LaMe} M. Landreman, B. Medasani and C. Zhu, \textit{Stellarator optimization for good magnetic surfaces at the same time as quasisymmetry}, Phys. Plasmas {\bf 28}, 092505 (2021).

\bibitem{Pl} G. G. Plunk, \textit{Perturbing an axisymmetric magnetic equilibrium to obtain a quasi-axisymmetric 
stellarator}, J. Plasma Phys.  {\bf 86}, 905860409 (2020).

\bibitem{Ro} E. Rodríguez, P. Helander, and A. Bhattacharjee 
 \textit{Necessary and sufficient conditions for quasisymmetry}, Phys. Plasmas {\bf 27}, 062501 (2020).

\bibitem{RoBa} E. Rodríguez and A. Bhattacharjee, \textit{Connection between quasisymmetric magnetic fields and anisotropic pressure equilibria in fusion plasmas}, Phys. Rev. E {\bf 104}, 015213 (2021).

\bibitem{CoPa} P. Constantin and F. Pasqualotto, \textit{Magnetic relaxation of a Voigt-MHD system}, Communications in Mathematical Physics {\bf 402}, 1931 (2023).

\bibitem{CaDu} R. Cardona, N. Duignan, and D. Perrella, \textit{Asymmetry of MHD equilibria for generic adapted
metrics},  Archive for Rational Mechanics and Analysis {\bf 249}, 1 (2025).

\bibitem{SoIl} E. A. Sorokina and V. I. Ilgisonis, \textit{Existence of true plasma equilibria in asymmetric magnetic fields}, Phys. Rev. E {\bf 110}, 065209 (2024).

\bibitem{So} E. A. Sorokina, \textit{Tokamak plasma equilibria with $n=1$ toroidal asymmetry}, Phys. Plasmas {\bf 31}, 112504 (2024).

\bibitem{IvSo} I. V. Ivanov and E. A. Sorokina, \textit{Plasma equilibrium configuration of Hill’s vortex type with $n = 3$ toroidal asymmetry}, Plasma Physics Reports  {\bf 51},  887 (2025).

\bibitem{LoSp} D. Lortz and G. O. Spies, \textit{Anisotropic magnetohydrostatic equilibria with arbitrary magnetic fields}, Phys. Plasmas  {\bf 10},  553 (2003).

\bibitem{Sol} L. S. Solov'ev, \textit{The theory of hydrodynamic stability of toroidal plasma configurations}, Soviet Physics JETP  {\bf 26},  400 (1968).

\bibitem{He} F. Herrnegger, Proceedings of V European Conference on Controlled Fusion and Plasma Physics, Vol. {\bf I}, 26 (1972).

\bibitem{Ma} E. K. Maschke, {\textit Exact solutions of the MHD equilibrium equation for a toroidal plasma}, Plasma Physics {\bf 15}, 535 (1973). 


\end{thebibliography}
\end{document}